
%
%
\magnification = \magstep1
\hsize = 15.5 true cm      
\vsize = 23.0 true cm      

\baselineskip  = 24.0 pt plus 1.0 pt minus 1.0 pt
\parindent     = 25.0 pt
\parskip       = 15.0 pt plus 5.0 pt minus 5.0 pt
  \hoffset -15.0 pt
%
%
\def \refs {\begingroup \frenchspacing
\parskip = 0.5 \baselineskip \parindent = 0 pt
\everypar = {\hangindent = 20.0 pt \hangafter = 1}}
\def \endrefs {\par \endgroup}

\def \draft
  {\def\today{\number\year
              /\ifnum\month<10 0\fi\number\month
              /\ifnum\day<10 0\fi\number\day}
   \headline={\hss\vrule\vbox{\hrule\smallskip
              \line {\hfil DRAFT\hfil \today \hfil}
              \smallskip\hrule}\vrule\hss}}

\font \small = cmr7

\vglue 2.0 cm

\centerline{\bf Giant Molecular Cloud Formation
through the Parker Instability}
\centerline{\bf in a Skewed Magnetic Field}

\bigskip

\centerline{T{\small OMOYUKI} H{\small ANAWA}}

\centerline{Department of Astrophysics, Nagoya University, Chikusa-ku,
Nagoya 464-01, Japan}

\bigskip

\centerline{R{\small YOJI} M{\small ATSUMOTO}}

\centerline{College of Arts and Sciences,
Chiba University, Yayoi-cho, Chiba 260, Japan}

\centerline{\small AND}

\centerline{K{\small AZUNARI} S{\small HIBATA}}

\centerline{National Astronomical Observatory, Mitaka, Tokyo 181, Japan}

\bigskip

\centerline{(Received 1991 December \ \ \ \ \ }

\bigskip

\centerline {\bf Abstract}

     The effect of the magnetic skew on the Parker instability is
investigated by means of the linear stability analysis for a gravitationally
stratified gas layer permeated by a horizontal magnetic field.
When the magnetic field is skewed (i.e., the field line
direction is a function of the height),
the wavelength of the most
unstable mode is $ \lambda \; \sim \; 10 H $ where $ H $ is the
pressure scale height.  The growth rate of the short wavelength modes
is greatly reduced
when the gradient in  the magnetic field direction exceeds
0.5 radian per scale height.  Our results indicate that
the Parker instability in a skewed magnetic field preferentially
forms large scale structures like
giant molecular clouds.

\noindent {\it Subject Headings}: hydromagnetics ----
instabilities ---- interstellar:magnetic fields.

\bigskip

\centerline{1. \ INTRODUCTION}

     The Parker instability has been thought to play a major role
on the formation of giant molecular clouds in galaxies
(Parker 1966; Shu 1974; Mouschovias 1974; Mouschovias, Shu, and Woodward
1974).  Although the Parker instability has many characteristics
favorable for the formation of giant molecular clouds, it
has one unfavorable characteristic in this regard:
the fastest growing mode of the Parker instability has
an infinite wave number perpendicular to the magnetic field line
and may produce many small gas condensations instead of
large scale gas condensations (Ass\'eo et al. 1978).
This argument has been used to suggest that the pure Parker
instability (without self-gravity)
leads to chaotic structures and turbulence rather than
large scale cloud complexes
(Ass\'eo et al. 1978; Elmegreen 1982).
This question, however, is based on a linear stability
analysis which assumes that the magnetic field is parallel in the
unperturbed state.  The Galactic magnetic field, however, is
not perfectly parallel (Heiles 1987; Sawa \& Fujimoto 1986; Shibata
\& Matsumoto 1991), but rather disturbed and skewed.
When the magnetic field is skewed, the mode having a large wave number
perpendicular to the magnetic field is likely to be suppressed, since
such modes force the adjacent field lines to intersect.
Thus, the fastest growing mode is expected to have a finite wave number
and to produce large scale condensations.
This {\it Letter} studies the effect of magnetic skew on the Parker
instability by means of linear stability analysis and supports
the above expectation.

     The stability analysis is carried out in \S 2 and its application
to giant molecular cloud formation is discussed in \S 3.

\bigskip

\centerline{2. \ LINEAR STABILITY ANALYSIS}

     We consider an idealized model of a skewed magnetic field
to evaluate the effect of the magnetic skew on the Parker
instability.  The model consists of an isothermal, stratified gas layer
permeated by a horizontal magnetic field whose direction
changes as a function of height at a constant pitch,
$ \varphi \, (z) \; \equiv \; \tan ^{-1} \; ( B _y / B _x ) \; \propto \; z $.
The density distribution and the magnetic field of
the model are
$$ \rho \; = \; \rho _0 \; {\rm exp} \; ( - z / H ) \; , \eqno (1) $$
$$ \eqalign{ {\bf B} \; &= \; ( B _x , \; B _y , \; B _z ) \cr
&= \; \lbrack B _0 \; \cos \;
( \varepsilon z / H ) \; {\rm exp} \; ( - z / 2H ) ,
\; B _0 \; \sin \; ( \varepsilon z / H ) \; {\rm exp} \; ( - z / 2H ) ,
\; 0 \rbrack \cr } \; , \eqno (2) $$
where $ H $ is the scale
height and $ \varepsilon $ denotes the strength
of the magnetic skew.
(The magnetic skew is not necessary to be constant
to suppress the growth of short wave length modes.
We imposed this assumption in order to minimize the number of
model parameters and to make our model as simple as possible.)
This model atmosphere is in hydrostatic
equilibrium under gravity in the $ z $-direction.
{}From the equation of motion, the scale height is expressed as
$$ g H \; = \; c _s {} ^2 \; + \; { B _0 {} ^2 \over 8 \pi \rho _0 } \; ,
\eqno (3) $$
where $ g $ is the gravitational acceleration and assumed to be
constant for simplicity.  The plasma beta,
$ \beta \; \equiv \; 8 \pi \rho c _s {} ^2 / B ^2 $,
is constant in this model atmosphere,
and is assumed to be unity, $ \beta \; = \; 1 $, in the following
except when otherwise is noted.
Figure 1 shows a schematic view of the
model magnetic field.  (Although this configuration of the magnetic
field is called \lq \lq magnetic shear\rq \rq \ in the solar physics,
we use the word \lq \lq skew\rq \rq \ in order to avoid the confusion
with the magnetic field sheared with the Galactic differential rotation.)

     We consider a small perturbation superimposed on the equilibrium
described above.  The perturbation is assumed to be isothermal and
to be described by the ideal MHD equations.
Since the perturbation equations are standard, their detailed form
is omitted here (see, e.g., Horiuchi et al. 1988
for perturbation equations).
A normal mode solution
of the perturbation equations has the form of
$$ \rho _1 (x, \; y, \; z) \; = \; \rho _1 (z) \; {\rm exp} \;
( - i \omega t \; + \; i k _x x \; + \; i k _y y ) \; . \eqno (4) $$
According to the variational principle (see, e.g., Bernstein 1983;
Nakamura, Hanawa, and Nakano 1991) the square of the eigenfrequency,
$ \omega ^2 $, is real.  The boundary condition is set so that
the energy density of the eigenfunction is periodic,
$$ \rho _1 (z \; + \; 2 \pi H / \varepsilon ) \; = \;
e ^{i \theta \; - \; \pi / \varepsilon } \;
\rho _1 (z) \; . \eqno (5) $$
See Shu (1974) and Nakamura, Hanawa, and Nakano (1991) for the reason
why the energy density should be constant in an eigenfunction.
The ratio of the phase shift, $ \theta $, to the vertical period
of the magnetic skew, $ 2 \pi H / \varepsilon $, is an effective
wavenumber in the vertical direction $
 ( k _{z, \, eff} \; = \; \theta \varepsilon / 2 \pi H ) $.
Since our model has the helical symmetry, the eigenfrequency
is a function of the total horizontal wave number,
$ \omega \; = \; \omega ( \sqrt{ k _x {} ^2 \; + \; k _y {} ^2 } ) $.
$ \lbrack $ Our model is invariant under the transformation of
$ (x, \; y, \; z ) \; \rightarrow \; ( x \cos \delta \; + \;
y \sin \delta , \; - \; x \sin \delta \; + \;
y \cos \delta , \; z \; + \; \delta H / \varepsilon ) $,
where $ \delta $ is an arbitrary number. $ \rbrack $

     Figure 2 shows the stability diagram of our model atmosphere
for $ \varepsilon \; = \; 1 $ and
$ \beta \; = \; 1 $.
The ordinate and the abscissa are the growth rate, $ - \omega ^2 H / g $,
and the wave number, $ k ^2 H ^2 $, respectively.  The growth rate and
the wave number are normalized by $ \sqrt{ g / H } $ and $ 1 / H $,
respectively.  The stability diagram has a band structure similar to
the energy level diagram of a crystal (cf. Kittel 1976).
The darkly painted regions denote the allowed bands and the blank
regions show the inhibited bands.
The most unstable mode has a finite wave number of $ k H \; = \; 0.70 $.

     On the top of the uppermost band $ v _{1z} $ of the eigenmode has
no node in the vertical direction $ ( \theta \, = \, 0 ) $.
The value of $ \theta $
changes from $ 2 n \pi $ to $ 2 (n \, + \, 1) \pi $ in each band.
As the number of nodes in the eigenfunction increases
($ \theta $ increases), the growth rate decreases continuously but
with some gaps.
This is a general property of the Parker instability.
When the initial magnetic field is parallel, the growth rate decreases
continuously as the wavenumber, $ \vert k _z \vert $, increases
(see, e.g., Parker 1979).  When the Parker unstable layer has a finite
thickness, the growth rate is discrete and decreases with the increase
in the number of nodes in the eigenfunction (Horiuchi et al. 1988).
In our model, the Parker unstable layers appear periodically in
the vertical direction for a given $ {\bf k} $.
In the band structure of figure 2 the growth rate is continuous in each
band and discrete between bands.  This is because our model atmosphere
has a periodical structure where each Parker unstable layer has a
finite thickness but the number of the Parker unstable layers is
infinite.

     Figure 3 shows the maximum growth at a given wave number for
$ \varepsilon \; = \; 0.2 $, 0.5, and 1.0.  The growth rate decreases
as $ \varepsilon $ increases for a given $ {\bf k} $.
The decrease is larger for a larger
wave number.  When $ \varepsilon \; \ge \; 0.2 $, the mode of
$ k \; = \; \infty $
is not the fastest growing mode.  The most unstable
mode has a wave length of $ \lambda \; \sim \; 10 H $ for
$ 0.2 \; \le \; \varepsilon \; \le \; 1.0 $.

    Figure 4 shows the maximum growth at $ k \; = \; \infty $
as a function of $ \varepsilon $.  As $ \varepsilon $ increases,
the growth rate at $ k \; = \; \infty $ decreases.
The growth of the mode having $ {\bf k} $ is suppressed by
magnetic tension when $ {\bf k} \cdot {\bf B} $ is large.
Whenever the magnetic field is not parallel and $ {\bf k} $ is
sufficiently large,
most magnetic field lines are bent with a short wavelength
and the magnetic tension suppresses the growth of the mode.
{}From figures 3 and 4 we find that the magnetic skew suppresses the
Parker instability of short wavelength modes and that the effect is
appreciable for $ \varepsilon \; \ge \; 0.5 $.
Note that $ \varepsilon \; = \; 0.5 $ corresponds to the case
when the direction of the magnetic field changes $ 29 ^\circ $ per
scale height.

\bigskip

\centerline{3. \ APPLICATION TO GIANT MOLECULAR CLOUD FORMATION}

     The analysis of the previous section demonstrated that
the fastest growing Parker instability has a wavelength of
$ \lambda \; = \; 10 H $ when the magnetic field is skewed
appreciably.  It implies that large scale mass condensations
are formed by the Parker instability if the Galactic magnetic field
is skewed significantly, i.e., more than several tens degrees per
scale height.  The Galactic magnetic field is not uniform in
the Solar neighborhood (see, e.g., Heiles 1987
and the references therein) and has a substantial nonuniform component.
Galactic dynamo theory (see, e.g., Sawa and Fujimoto 1986;
Fujimoto and Sawa 1987) also predicts
that the
Galactic magnetic field changes its direction as a function of
the height; in their model the direction of the magnetic field
in the halo (at the height of $ z \; = \; 1 \; {\rm kpc} \; \sim
\; 6 H $)
differs $ 180 ^\circ $ from that in the disk
$ ( \varepsilon \; \sim \; 0.5 ) $.  Also the nonlinear development of the
Parker instability produces the skewed
magnetic field owing to the Coriolis force, even when the initial field
is not skewed (Shibata and Matsumoto 1991).  Consequently, although
a firm observational evidence of the magnetic skew has not
been obtained, it is very likely that the Galactic magnetic field has a
skew component which is significantly  strong for the formation
of giant molecular clouds through the Parker instability.

     Our results (Fig. 3) indicate that the growth time of
the Parker instability is
$$ \tau \; \simeq \; 4.9 \times 10 ^7 \;
\Bigl( { \omega ^2 H / g \over 5.8 \times 10 ^{-2} } \Bigr) ^{-1/2} \;
\Bigl( { H \over 160 \; {\rm pc} } \Bigr) ^{1/2} \;
\Bigl( { g \over 3.5 \times 10 ^{-9} \; {\rm cm} \; {\rm s} ^{-2} }
\Bigr) ^{-1/2} \; {\rm years} \; , \eqno (6) $$
in a skewed field with $ \varepsilon \; \simeq \;
0.2 - 1.0 $ and $ \beta \; = \; P _g / P _m \; = \; 1 $
and the mass of the cloud formed by the Parker instability is
$$ M \; \sim \; 2 \rho \lambda ^2 H \; \sim \; 2 \times 10 ^7 \;
\Bigl( { n \over 1 \; {\rm cm} ^{-3} } \Bigr) \;
\Bigl( { \lambda \over 10 H } \Bigr) ^2 \;
\Bigl( { H \over 160 \; {\rm pc} } \Bigr) ^3 \; {\rm M} _\odot \; .
\eqno (7) $$

     We now briefly comment on the effect of the self-gravity on
the Parker instability.  Hanawa, Nakamura, and Nakano (1991) has
analyzed the linear stability of the Galactic gaseous disk taking
account of the magnetic field, the Galactic rotation, and
the self-gravity of the gas.   The magnetic field was assumed to
be parallel in the equilibrium.  According to them, the growth rate
is larger when the wavenumber perpendicular to the field line is larger.
The growth of short wavelength modes are not suppressed by the
self-gravity. \lq \lq Turbulence\rq \rq \ rather than a cloud is
likely to be formed as far as the magnetic field is parallel
in equilibrium even if the self-gravity is taken into account.
Thus, the skewed magnetic field is an important factor to produce
a large scale structure like clouds through the Parker instability.

     The skewed magnetic field may affect not only the linear growth
rate of the Parker instability but also the nonlinear evolution
of the Parker unstable modes.  The intersection of the magnetic
field lines is strictly inhibited in the framework of the ideal
magnetohydrodynamics.  Thus, the growth of an unstable
mode will be saturated at latest at the stage when  simple extrapolation
of the linear stability analysis predicts the intersection of the magnetic
field lines (see Matsumoto et al 1988, 1990 for justification
of this conjecture).   Figure 5 shows the eigenfunction of
the Parker unstable mode in the skewed magnetic field
$ ( \varepsilon \; = \; 0.5 \; {\rm and} \; \beta \; = \; 1) $.
The ordinate is the velocity perturbation, $ v _{1z} $, at the
origin $ ( \; x , \; y ) \; = \; ( \; 0 , \; 0 ) $ and the
abscissa is the height $ z $.
The thick curve is for the mode of $ ( k _x , \; k _y ) \; = \;
( \; 0.7 / H , \; 0 ) $ and the thin curve for that of
$ ( k _x , \; k _y ) \; = \; ( \; 0 , \; 0.7 / H ) $.
As shown in the figure the velocity, $ v _{1,z} $, is not a monotonically
increasing function of $ z $.  At some height, the rising magnetic
field line catches up with the foregoing field line as the perturbation
grows.  The growth of the perturbation is likely to be saturated
as the field lines becomes closer to each other.
If the two eigenmodes shown in figure 5 are excited simultaneously,
however,
the field lines are less likely to catch up with the neighboring
field lines.  The linear combination of the eigenmodes is also an
eigenmode since the growth rates are the same
$ \lbrack \omega ( {\bf k} ) \; = \; \omega ( \vert {\bf k} \vert )
\rbrack $.
It implies that the perturbation is less easy to be saturated
when more than 2 plane waves are simultaneously excited.
If all the mode having the same wave number, $ \vert {\bf k} \vert $,
are excited, the perturbation velocity increases with height monotonically
$ \lbrack v _{1z} \; \propto \; \rho _0 {} ^{-1/2} \; \propto \;
{\rm exp} \; ( z / 2 H ) \rbrack $
at $ ( \; x , \; y ) \; = \; ( \; 0 , \; 0 ) $.
Then, the perturbation grows up to have a large amplitude without
suffering serious nonlinear effects.  The linear combination of
many plane waves leads to a cylindrical wave having a symmetry axis
at $ ( \; x , \; y ) \; = \; ( \; 0 , \; 0 ) $ (see Fig. 5).
Thus, we expect that
a circular structure would be formed in a skewed magnetic
field rather than a stripe.

      The authors thank Professors M. Fujimoto and T. Sawa for
useful discussions and encouragement, and Dr. A.~C. Sterling for
useful comments.  The numerical computation
was performed mainly at The Computer Center at Nagoya University.
The present work is financially supported in part by the Grant-in-Aid
for Cooperative Scientific Research by the Ministry of Education,
Science, and Culture of Japan (03302013).

\bigskip

\vfill \eject

\noindent {\bf References}

\refs

Ass\'eo, E., Cesarsky, C.~J., Lachieze-Rey, M., \& Pellat, R.
     1978, ApJ, 225, L21.

Bernstein, I.~B. 1983, in Handbook of Plasma Physics Vol. 1:
     Basic Plasma Physics I, ed.
     A.~A. Galeev and R.~N. Sudan (Amsterdam: North-Holland Publishing),
     Chapter 3.1.

Elmegreen, B. 1982, ApJ, 253, 634.

Fujimoto, M. \& Sawa, T. 1987, PASJ, 39, 375.

Heiles, C. 1987, in Interstellar Processes, ed. D.~.J. Hollenbach
     \& H.~A. Thronson, Jr. (Dordrecht: Reidel), 171.

Horiuchi, T., Matsumoto, R., Hanawa, T., \& Shibata, K. 1988,
     PASJ, 40, 147.

Kittel, C. 1976, Introduction to Solid State Physics 5th ed. (New York:
     Wiley), 183.

Matsumoto, R., Horiuchi, T., Shibata, K., \& Hanawa, T. 1988,
     PASJ, 40, 171.

atsumoto, R., Horiuchi, T., Hanawa, T., \& Shibata, K. 1990,
     ApJ, 356, 259.

Mouschovias, T.~Ch. 1974, ApJ, 192, 37.

Mouschovias, T.~Ch., Shu, F.~H., \& Woodward, P.~R. 1974, A\&Ap, 33, 73.

Nakamura, F., Hanawa, T., \& Nakano, T. 1991, PASJ, 43, 685.

Parker, E.~N. 1966, ApJ, 145, 811.

Parker, E.~N. 1979, Cosmical Magnetic Fields
     (Oxford University Press, Oxford), p. 314.

Sawa, T. \& Fujimoto, M. 1986, PASJ, 38, 133.

Shibata, K. \& Matsumoto, R. 1991, Nature, 353, 633.

Shu, F.~H. 1974, A\&Ap, 33, 55

\endrefs

\vfill \eject

\vfill \eject

\centerline{\bf Figure Legends}

\item{\bf Fig. 1.} Schematic view of the skewed magnetic field.

\item{\bf Fig. 2.} The stability diagram for $ \varepsilon \; = \; 0.5 $
and $ \beta \; = \; 1 $.
The ordinate is the square of the growth rate,
$ - ( H / g ) \omega ^2 $, and the abscissa is the square of the
wave number, $ H ^2 k ^2 $.  The eigenmode has a band structure.
The allowed bands are denoted with black and the inhibited bands
are denoted with blank.

\item{\bf Fig. 3.}  The maximum growth rate at a given wave number for
$ \varepsilon \; = \; 0.2 $, 0.5, and 1.0.

\item{\bf Fig. 4.}  The maximum growth rate at $ k \; = \; \infty $
as a function of $ \varepsilon $.

\item{\bf Fig. 5.}  The velocity perturbation, $ v _{1z} $,
of the unstable eigenmode at $ ( \; x , \; y ) \; = \; ( \; 0 , \; 0 ) $
as a function of the height, $ z $
for $ \varepsilon \; = \; 0.5 $, $ \beta \; = \; 1.0 $,
$ \vert {\bf k} \vert H \; = \; 0.7 $,
and $ \omega ^2 H / g \; = \; - 5.82 \times 10 ^{-2} $.
The thick solid curve is for the mode of $ ( \; k _x H , \; k _y H ) \; =
\; ( \; 0.7 , \; 0.0 ) $ and the thin solid curve for that of
$ ( \; k _x H, \; k _y H) \;  = \; ( \; 0.0 , \; 0.7 ) $.
The dashed curve denotes the linear combination of many plane wave
eigenmodes, where the velocity perturbation is proportional to
$ v _{1z} \; \propto \; \rho ^{-1/2} $ at
$ ( \; x , \; y ) \; = \; ( \; 0 , \; 0 ) $.

\bye